\documentclass[reprint,aps,prb,twocolumn,superscriptaddress,showpacs,longbibliography]{revtex4-1}

\usepackage[pdftex]{graphicx} \usepackage{float}

\usepackage{amsmath} \usepackage{amsfonts} \usepackage{amssymb}

\usepackage[pdftex,colorlinks=true]{hyperref}
\usepackage{mathtools} 
\DeclarePairedDelimiter\ket{\lvert}{\rangle}
\DeclarePairedDelimiterX\braket[2]{\langle}{\rangle}{#1 \delimsize\vert #2}
\DeclarePairedDelimiterX\expval[3]{\langle}{\rangle}{#1 \delimsize\vert #2
  \delimsize\vert #3}

\usepackage{mathtools}

\newcommand{\vect}[1]{\mathbf{#1}}

\def\sectionn#1{\noindent\underline{\it #1:}}

\begin{document}

\title{Disordered flat bands on the kagome lattice}

\author{Thomas Bilitewski} \affiliation{Max-Planck-Institut f\"{u}r Physik
  komplexer Systeme, N\"othnitzer Str.\ 38, 01187 Dresden, Germany}

\author{Roderich Moessner} \affiliation{Max-Planck-Institut f\"{u}r Physik
  komplexer Systeme, N\"othnitzer Str.\ 38, 01187 Dresden, Germany}

\begin{abstract}
  We study two models of correlated bond- and site-disorder on the kagome
  lattice considering both translationally invariant and completely disordered
  systems. The models are shown to exhibit a perfectly flat ground state band in
  the presence of disorder for which we provide exact analytic solutions.
  Whereas in one model the flat band remains gapped and touches the dispersive
  band, the other model has a finite gap, demonstrating that the band touching
  is not protected by topology alone. Our model also displays fully saturated
  ferromagnetic groundstates in the presence of repulsive interactions, an
  example of disordered flat band ferromagnetism.
\end{abstract}

\maketitle

\section{Introduction} The physics of flat bands has generated considerable
excitement over the years \cite{Tasaki1998,Derzhko2015,Leykam2018} In a flat
band, the kinetic energy is completely suppressed; thus, transport is hindered
by a vanishing group velocity, and any kind of interaction is non-perturbative
in nature and can mix the extensive number of degenerate states in the flat
band, with the potential to create complex many-body states and phenomena. One
well known example of this mechanism at work is the fractional quantum Hall
effect, where interactions induce highly non-trivial behaviour of the electrons
in the degenerate Landau levels of a magnetic field.

Thus, flat band systems are well-suited for producing unconventional phenomena
\cite{Parameswaran2013,BERGHOLTZ2013, Derzhko2015}. For both fermions and
bosons, they allow to realise the fractional quantum hall effect in absence of a
magnetic field \cite{Sheng2011,Wang2011,Neupert2011,Sun2011}, i.e. fractional
Chern Insulators, and at potentially high temperatures \cite{Tang2011}. Other
contexts include high-temperature superconductivity \cite{Imada2000,Peotta2015},
Wigner crystalisation \cite{Wu2007,Jaworowski2018}, realising higher-spin
analogs of Weyl-fermions \cite{Dora2011}, bands with chiral character
\cite{Ramachandran2017}, lattice super-solids \cite{Huber2010}, fractal
geometries \cite{Pal2018}, magnets with dipolar-interactions
\cite{Maksymenko2017}, and Floquet physics \cite{Du2017,Roman-Taboada2017}. Flat
bands of magnons also play a crucial role in determining the behaviour of
quantum magnets in magnetic fields
\cite{Schulenburg2002,Zhitomirsky2005,Schmidt2006,Derzhko2007}.

Interest in flat band physics is not restricted to the presence of interactions,
but also extends to their response to disorder, as the flat band states can turn
out to be critical displaying multifractality \cite{Chalker2010}, or
unconventional localisation behaviour \cite{Flach2014,Bodyfelt2014,Leykam2017}.
They also appear in purely classical mechanical systems \cite{Perchikov2017},
and in the field of photonics \cite{Zong2016,Leykam2018a}. Quite recently, flat
bands have been experimentally demonstrated in a realistic Kagome material
\cite{Lin2018} as well as in optical lattices \cite{Jo2012}.

\begin{figure}
  \begin{minipage}{0.99\columnwidth}
    \includegraphics[width=.99\columnwidth]{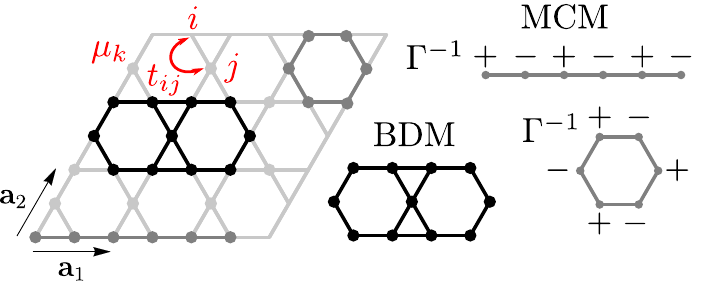}
  \end{minipage}
 \caption{Kagome lattice with lattice vectors $a_1$ and $a_2$, shown is a
   finite-size lattice with $L_x=L_y=3$, opposite edges are identified for
   periodic boundary conditions. The model contains site-dependent nearest-neighbour tunnelings
   $t_{ij}$ and chemical potentials $\mu_k$.
   The highlighted sites correspond to a zero-energy flat band-state of the MCM, hexagon
   and system-spanning loop (dark gray) or the BDM, double hexagon (black).
   \label{fig:illustration_lattice}}
\end{figure}

In this work consider non-interacting nearest neighbour hopping models on the
Kagome lattice with correlated bond- and site-disoder, as illustrated in
Fig.~\ref{fig:illustration_lattice}. The simple nearest neighbour hopping model
on the Kagome lattice is known to host a degenerate flat band
\cite{Mielke1991,Mielke1991a,Mielke1992,Tasaki1992,Mielke1993} with a quadratic
band touching point believed to be topologically protected\cite{Bergman2008}.
However, in interacting many-body physics it is often preferable to work with a
gapped flat band to protect it from 'Landau-level mixing', i.e.\ from
interactions with the dispersive bands.

Here, we explicitly construct a gapped flat band on the Kagome lattice. The
simplest setting in which it appears contains modulated bond and site-disorder,
both in presence of translational symmetry (where one can speak of a band) and
in absence of it, i.e. in the presence of random disorder, where one may still
identify an extensive manifold of degenerate states. In fact, we find that a
local perturbation to the Hamiltonian can open a gap above the flat band. This
indicates that the band-touching is protected not just by topology but requires
also symmetry.

We obtain exact solutions for the flat band states of all of these models,
facilitating a clear interpretation of why the chosen type of correlated
site-bond-disorder does not lift the extensive degeneracy of the flat band, and
providing new insight into the stability of the flat bands and the protection of
the quadratic band-touching point. Our study also adds an example where
compactly localised Wannier-states can be explicitly constructed for a
disordered flat band model.

Our treatment extends previous observations on the flat band in kagome, such as
the observed stability of the flat band and band-touching points to breathing
anisotropy \cite{Essafi2017}, and opens up new perspectives: We show how to
selectively gap out the flat band, or the Dirac cones, or all bands. Thus, our
results reinforces the role of the kagome lattice as a platform for the study of
topological physics and flat band physics in general, in particular the physics
of perturbations and disorder in flat bands.

\section{Model}
We study non-interacting particles on the kagome lattice
\begin{equation}
  \mathcal{H} =  \sum_{\left<i,j\right>} \left(  t_{ij} \hat{c}^{\dagger}_{i} \hat{c}_j +c.c. \right)+  \sum_i \mu_i \hat{n}_i \, ,
\end{equation}
with nearest-neighbour (complex) hoppings $t_{ij}$ between sites $i,j$ and
site-dependent chemical potentials $\mu_i $ at site $i$. In the models we
consider $\mu_i$ is given as a function of the couplings $t_{ij}$. The specific
correlation between the hopping and potential terms is motivated by a connection
to bond-disordered Heisenberg models \cite{Bilitewski2017} where it naturally
arises via an exact rewriting of the Hamiltonian.

The Hamiltonian can be compactly written via its matrix elements $H_{ij}$ as
$\mathcal{H} = \sum_{ij} c_i^{\dagger} H_{ij} c_j$. Noting that this only
connects nearest neighbours, and that every nearest neighbour pair belongs
either to an up or down triangle of the Kagome lattice, we rewrite the
Hamiltonian in the following way
\begin{align}
  \mathcal{H} &= \mathcal{H}^{\vartriangle} + \mathcal{H}^{\triangledown} \\
  H_{ij}^{\vartriangle / \triangledown} &=\begin{cases}  \bar{\gamma}^{\vartriangle / \triangledown}_i \gamma^{\vartriangle / \triangledown}_j + |\gamma^{\vartriangle / \triangledown}_i|^2 \delta_{ij} , & \text{for } i,j \in \alpha \\
    0  , &\text{otherwise} \end{cases}
           \label{eq:H_split}
\end{align}
where we first split it into its contribution on the up and down triangles, and
then define all couplings within a triangle $\alpha$ via site and triangle
dependent (complex) factors $\gamma^{\vartriangle/\triangledown}_{i}$.

This form makes the correlation between the hoppings and chemical potentials
explicit. Specifically, we have $t_{ij} = \bar{\gamma}_i^{\alpha}
\gamma_j^{\alpha}$ for sites $i,j$ in the triangle $\alpha$ and $\mu_i =
|\gamma_i^{\vartriangle}|^2+|\gamma_i^{\triangledown}|^2$. In the presence of
lattice-inversion symmetry $\mathcal{H}^{\vartriangle} =
\mathcal{H}^{\triangledown}$ and these factors become solely site-dependent. We
will refer to the model with lattice inversion symmetry as the maximal Coulomb
model (MCM), and with broken lattice inversion symmetry as the bond-disordered
model (BDM).

This also allows us to make an insightful connection to the Hamiltonian of the
non-disordered model, essentially the disordered model can be understood as a
rescaling of the clean model by the $\gamma$ factors. Using that the Hamiltonian
is fully specified by its matrix elements $H_{ij}$, we can further split them as
a product of three matrices as
\begin{equation}
  H^{\vartriangle/\triangledown} = \bar{\Gamma}^{\vartriangle/\triangledown} H^{\vartriangle/\triangledown}_0 \Gamma^{\vartriangle/ \triangledown}
  \label{eq:H_factors}
\end{equation}
with $\Gamma^{\vartriangle/\triangledown}_{ij} = \delta_{ij}
\gamma_i^{\vartriangle/\triangledown}$, a diagonal matrix containing the scaling
factors, and $H_0$ the matrix of the clean system with $\gamma_i^{\alpha} \equiv
1$, describing the nearest neighbour hopping on the kagome lattice.

Making use of the form $\mathcal{H}=\sum_{ij} c^{\dagger}_i H_{ij} c_j$ the
action of the Hamiltonian on single particle states $\ket{\Psi}=\sum_i \psi_i
c_i^{\dagger} \ket{\mathrm{vac}}$ is simply
\begin{equation}
  \mathcal{H} \ket{\Psi} = \sum_i H_{ik} \psi_k c^{\dagger}_i\ket{\mathrm{vac}}= \sum_i (H \psi)_i c^{\dagger}_i \ket{\mathrm{vac}} \, .
  \label{eq:H_action}
\end{equation}
From this we obtain the expectation value as
\begin{equation}
  \expval{\Psi}{\mathcal{H}}{\Psi} = \sum_{ij} \bar{\psi}_i H_{ij} \psi_j = \sum_{\alpha}  \left|  \sum_{i\in\alpha}\gamma_{i}^{\alpha}\psi_i \right|^2  =
  \sum_{\alpha}  \left|\psi_{\alpha} \right|^2 ,
\end{equation}
where in the second equality we used the explicit form of the Hamiltonian,
Eq.~\ref{eq:H_split}, which splits into a sum over triangles $\alpha$, and in
the last equality defined the sum of scaled amplitudes within a triangle
$\psi_{\alpha} = \sum_{i \in \alpha} \gamma_{i}^{\alpha} \psi_i$.

Thus, exact zero-modes are states with $\psi_{\alpha}=0$ on all triangles
$\alpha$. This condition is typically referred to as a groundstate constraint in
the theory of frustrated magnets and is intimately connected to height-mappings
and emergent gauge theory descriptions of the groundstate phase. For spins the
condition $\psi_{\alpha}=0$ is more stringent and can only be fulfilled for not
too disparate bond values due to the unit length constraint which is found to
lead to a phase transition of the model. In contrast, here it can be fulfilled
for arbitrary choices.

\section{Construction of Flat Band states}
\sectionn{Exact Mapping of flat band for the MCM} The clean system is known to
host an exactly flat band at $E=0$ which touches the dispersive band at $q=0$
\cite{Bergman2008}.

In the non-disordered model ($\gamma_{i}^{\alpha}=1$), the ground state
condition $\psi_{\alpha} = \sum_{i \in \alpha} \psi_i=0$ reduces to the simple
sum of amplitudes in every triangle vanishing. It is easy to check that the
states illustrated in Fig.~\ref{fig:illustration_lattice}, a hexagon loop with
alternating $+,-$, and a system-spanning loop with alternating $+,-$ amplitudes,
satisfy this, and (less-trivially) that these yield $N_s/3+1$ linearly
independent zero-energy states. Since the kagome lattice has 3 sites in the unit
cell and thus 3 bands, finding $N_s/3+1$ states at the same energy then also implies
the band-touching.

For the MCM all these zero-modes of the clean system can be mapped
to zero-modes of the disordered model via
\begin{equation}
  \Psi^{\mathrm{FB}}_{\mathrm{MCM}} = \Gamma^{-1} \Psi^{\mathrm{FB}}_{0} \, ,
\end{equation}
which follows directly from $H^{\vartriangle} = H^{\triangledown}$ in the MCM
together with Eq.~\ref{eq:H_factors} and Eq.~\ref{eq:H_action}, e.g. the
observation that the disordered model can be understood as a rescaling of the
clean model. Thus, we obtain an exactly flat band at $E=0$. This
further implies that the band touching point is preserved as well.

The flat band states of the MCM can therefore be characterised the same way as
in the clean system \cite{Bergman2008}: The MCM (a) $N_s/3+1$ zero-modes, (b) of
which $(N_s/3-1)$ can be chosen as lin. independent localised hexagon loop modes
and 2 as system-spanning delocalised loops (both types arising via the mapping
from the zero-modes of the clean system), and (c) the flat band is gapless
touching the dispersive band. The two different types of states are
schematically illustrated in Fig.~\ref{fig:illustration_lattice}.

We emphasise that this is completely independent of the specifics of $\gamma_i$,
e.g. it holds true for translationally invariant, completely disordered,
positive, negative and sign-changing, and real or complex choices. In fact, it
holds true for a slightly more general model, where $\Gamma^{\vartriangle} = c
\, \Gamma^{\triangledown} $ which in particular includes the model with
breathing anisotropy.

\sectionn{Construction of flat band for BDM}
\begin{figure}
  \begin{minipage}{0.99\columnwidth}
    \includegraphics[width=.99\columnwidth]{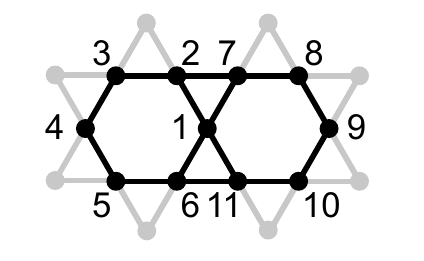}
  \end{minipage}
 \caption{A double hexagon of the kagome lattice. The wavefunction of a BDM zero
   energy state is localised on the black sites.
   Note that the state occupies 11 sites, and is part of 10 triangles, thus,
there are 11 degrees of freedom and 10 constraints, in addition to the
   wavefunction normalisation, implying that there is a unique solution for such
   a localised state.
   \label{fig:illustration_double_hexagon}}
\end{figure}
We note that such a mapping is not possible for the BDM where $\Gamma$ differs
non-trivially between up- and down-triangles. Thus, it is not immediately
obvious that the BDM should host an extensively degenerate groundstate band and
if so whether the band-touching point is preserved.

We first summarise the findings and then provide a construction of the flat band
states. We find that (a) the BDM has $N_{s}/3$ exact zero-modes/flat band, (b)
the flat band states states can all be localised and (c) the flat band is
generically gapped.

We emphasise the last point, stating that it is possible to maintain the
flatness of the band while gapping it from the dispersive bands in contrast to
the claimed topological protection \cite{Bergman2008}. We will analytically show
this in the next section for translationally invariant model, and provide
numerical evidence for disordered systems. In fact, it is sufficient to break
inversion symmetry by changing a single coupling $\gamma_{i}^{\vartriangle}$ to
create a gap to the flat band.

We now explicitly construct the $N_s/3$ linearly independent localised states
forming the degenerate flat band. To do so, we consider a double hexagon of the
kagome lattice shown with our conventions for the site labels in
Fig.~\ref{fig:illustration_double_hexagon}. We note that such a state occupies
11 sites and these sites are part of 10 triangles of the kagome lattice. Each
triangle contributes one scalar constraint $\Psi_{\alpha}=0$, in addition to one
normalisation constraint, thus, we might expect a unique solution on every
hexagon-pair.

The resulting linear system of equations can be solved explicitly (see SM
\cite{supplemental}), and the wave-function amplitudes may be written as a
function of the coupling terms $\gamma_i^{\alpha}$ as $\Psi_i = \Psi_1 \,
f_i(\gamma_i^{\alpha})/D(\gamma_i^{\alpha}) $. This solution is only valid if
the determinant $D$ given by
\begin{equation}
  \Delta = \gamma^{\triangledown}_3 \gamma^{\triangledown}_5 \gamma^{\triangledown}_7 \gamma^{\triangledown}_9 \gamma^{\triangledown}_{11} \,
  \gamma^{\vartriangle}_2 \gamma^{\vartriangle}_4 \gamma^{\vartriangle}_6 \gamma^{\vartriangle}_8 \gamma^{\vartriangle}_{10} - (\triangledown \leftrightarrow \vartriangle) \, ,
\end{equation}
is non-zero. This manifestly vanishes in the presence of inversion symmetry
($\gamma^{\vartriangle} =\gamma^{\triangledown} $), but is non-zero if inversion
symmetry is broken ($\gamma^{\vartriangle} \neq \gamma^{\triangledown} $).
Therefore, in the BDM there is a unique localised state on every double-hexagon.

We have checked (numerically) that taking $L^2$ such double-hexagons tiling the
full kagome lattice does yield $L^2$ independent states, thus, providing a full
basis for the zero-energy states of the BDM, in contrast to the MCM and the
clean system which requires the system spanning loop states \cite{Bergman2008}.

It is also easy to show that no such solution for a localised state is possible
on a single heaxagon (see SM \cite{supplemental}), thus, proving that these
found states indeed form a maximally localised basis of the flat band manifold.

Typically, in presence of interactions the size of the maximally localised basis
states strongly affects the behaviour of the model, and here we find that this
size doubles in presence of infinitesimal disorder. In fact, the existence of a
compactly localised basis for flat bands is an open question of research with
relations to the topology of the corresponding Bloch bands
\cite{Maimaiti2017,Maimaiti2018,Rhim2018}.

\section{Gapped flat bands}
It remains to show that the BDM flat band states are indeed gapped and do not
touch the dispersive bands, which we will show in the next sections both for
translationally invariant and generic disordered models.

\sectionn{Translationally invariant systems}
\begin{figure}
  \begin{minipage}{0.99\columnwidth}
    \includegraphics[width=.99\columnwidth]{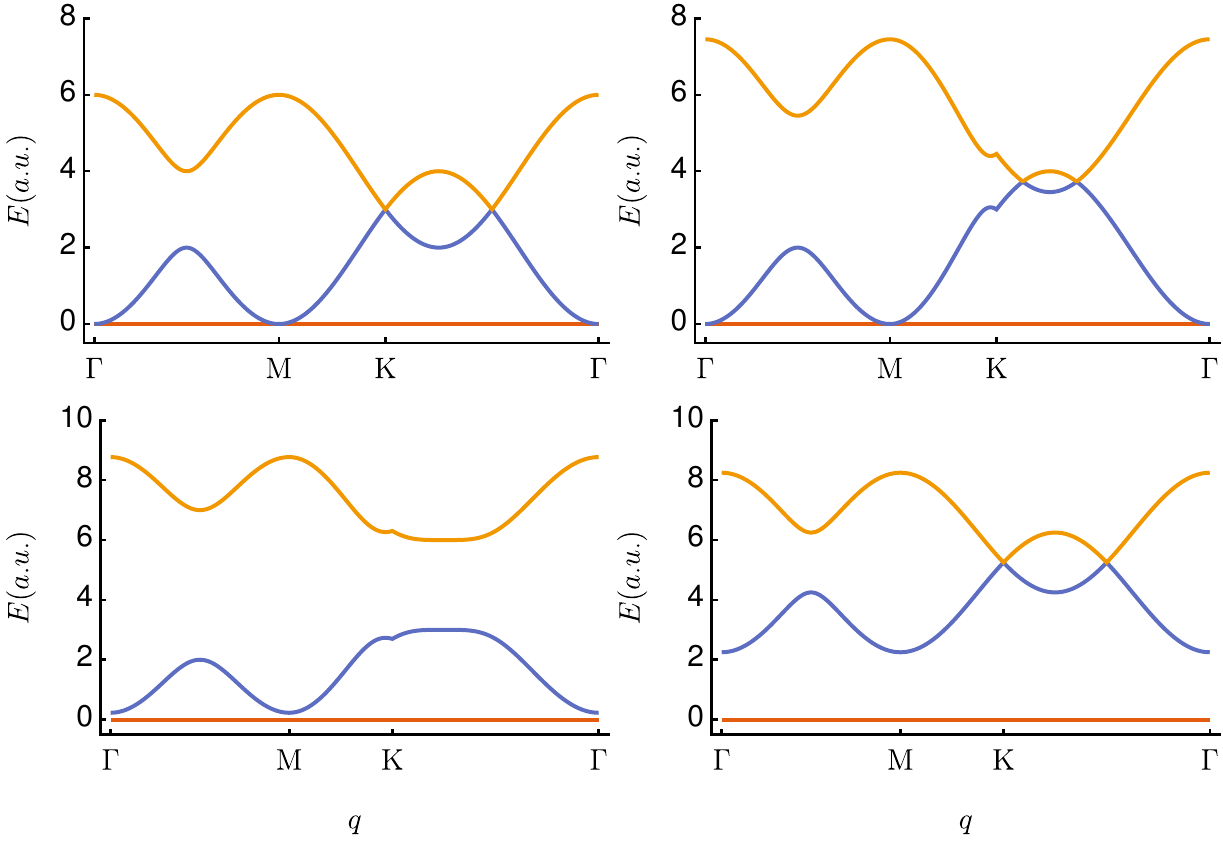}
  \end{minipage}
 \caption{ Dispersion along high-symmetry lines in the Brillouin zone.
   From top left to bottom right: clean system, MCM with 
   $\gamma_A=\gamma_A^{\vartriangle}=\gamma_A^{\triangledown}<\sqrt{2}$, BDM with
   $\gamma_A=2$ and BDM with $\gamma_A^{\vartriangle}= \frac{1}{\gamma_A^{\triangledown}}=\gamma_B^{\triangledown}=\frac{1}{\gamma_B^{\vartriangle}}= 0.5$.
   \label{fig:band_structure}}
\end{figure}
We begin by considering translationally invariant systems with real couplings.
In that case the model has 6 (3) free parameters
$\gamma_{A,B,C}^{\vartriangle/\triangledown}$ for BDM (MCM), e.g. the couplings
on the three sites (A,B,C) in a triangle of the Kagome lattice, with different
couplings on the up and down triangle for the BDM model.

In this case, one can analyse the model in momentum space, and analytical
results can be obtained (see SM \cite{supplemental}). We find that for every $q$
there is exactly one zero-mode, i.e. we find a flat band at $E=0$ for both the
BDM and MCM as anticipated from the construction of the zero-modes above.
Importantly, this allows us to obtain an analytic expression for the gap of the
BDM, thus, proving our claim that the BDM flat band can indeed be gapped.

We will consider illustrative examples for the gap below, please see SM
\cite{supplemental} for the general expression of the gap. As the simplest model
consider just $\gamma_A^{\vartriangle} \neq 1$, then the gap scales as
\begin{equation}
  \Delta_{\mathrm{gap}}=\frac{1}{2} \left(5+  {\gamma_A^{\vartriangle}}^2 -\sqrt{\left( {\gamma_A^{\vartriangle}}^2+1 \right)^2  +16 \gamma_A^{\vartriangle}+16}\right) 
\end{equation}
%
showing a quadratic scaling for small deviations away from the homogeneous
system.

A more symmetric arrangement can be obtained by considering
$\gamma_A^{\vartriangle} = \frac{1}{\gamma_A^{\triangledown}}
=\gamma_B^{\triangledown} =\frac{1}{\gamma_B^{\vartriangle}}=x$, which yields
the gap to the flat band as
\begin{equation}
  \Delta_{\mathrm{gap}} = x^2 +x^{-2} -2 \, .
\end{equation}
We note that this allows to cleanly separate the flat band by an (arbitrarily)
large gap from all dispersive bands, making the Kagome lattice a prime platform
to study physics in flat bands.

We show dispersion relations along high-symmetry lines in the Brillouin zone for
the clean model, the MCM and the BDM in Fig.~\ref{fig:band_structure}. We
emphasise that clearly both models retain an exactly flat band at $E=0$. As
discussed above the MCM always retains the band-touching point at $q=0$
($\Gamma$ point), but the Dirac-points can be gapped for large perturbations
(not shown).

In contrast in the BDM, the flat band is always gapped as seen in the lower
panel of Fig.~\ref{fig:band_structure}, already for infinitesimal changes in the
couplings. Just changing a single coupling generically gaps both the flat band
and the Dirac points (lower left panel). For the symmetric choice described
above, the flat band is gapped, but the Dirac points remain gapless (lower right
panel).

In summary, we have shown that we can selectively gap out the flat bands and
keep the Dirac cones or gap out the Dirac cones, but keep the quadratic band
touching point, or gap out all bands.

\sectionn{Local Perturbation} Before considering fully disordered models it is
insightful to understand the effect of a local perturbation to the system. For a
topologically protected band-crossing one would expect the resulting gap to
scale to zero exponentially in system size.

\begin{figure}
  \begin{minipage}{0.49\columnwidth}
    \includegraphics[width=.99\columnwidth]{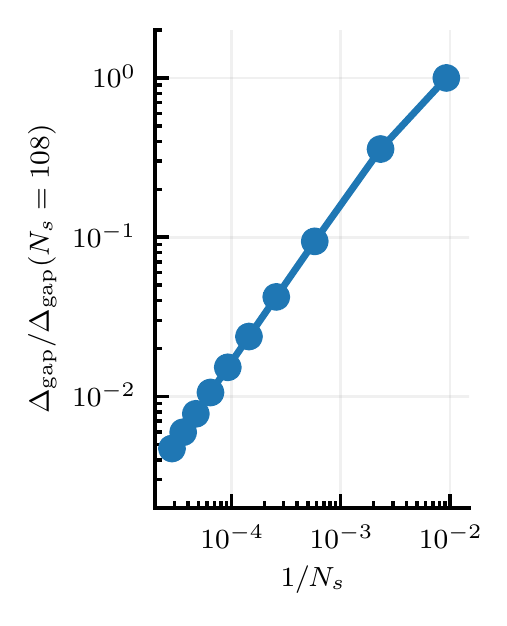}
  \end{minipage}
  \begin{minipage}{0.49\columnwidth}
    \includegraphics[width=.99\columnwidth]{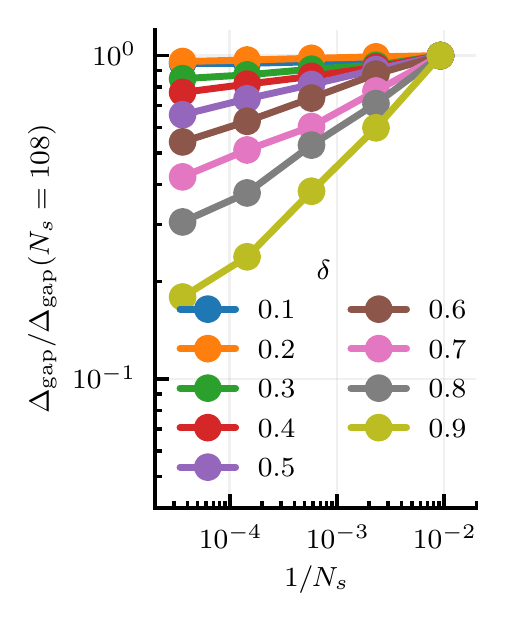}
  \end{minipage}
 \caption{(a) Gap of the flat band in presence of a local perturbation versus
   inverse number
   of sites $N_s$ on a log-log scale showing a linear scaling with inverse
   number of sites $\sim N_s^{-1}$.
   (b) Gap of the flat band for the fully disordered system versus inverse
   number of sites for different disorder strengths $\delta$.
   \label{fig:gaps}}
\end{figure}

We modify the Hamiltonian locally by changing a single coupling
$\gamma^{\vartriangle}$ affecting one site potential $\mu$ and two tunnel
couplings $t$. As a result, in Fig.~\ref{fig:gaps}(a) we observe a linear
decrease of the gap with inverse number of sites $\sim N_s^{-1}$, consistent
with the gap closing in the thermodynamic limit. However, the decay is clearly
not exponential as would be expected for a topologically protected degeneracy.

\sectionn{Disordered Systems} Next, we consider fully disordered models with
random choices for $\gamma^{\vartriangle/\triangledown}_{i}$. As an example we
consider a box-uniform distribution $\gamma \in [1-\delta,1+\delta]$. However,
we emphasise that this specific choice is not relevant and the conclusions hold
true for any generic disorder distribution.

The gap to the flat band versus inverse system size for a range of values of
$\delta$ is shown in Fig.~\ref{fig:gaps}(b). It extrapolates to a finite value
in the thermodynamic limit for $\delta < 1$, and scales as $\delta^2$ for small
disorder strengths. Thus, we conclude that disorder of this type gaps out the
flat band, even for infinitesimal disorder strength.

We also note in passing that the finite gap implies that the projector into the
flat band decays exponentially for the BDM model, but decays algebraically for
the gapless MCM.

\sectionn{Flat Band Ferromagnetism in a disordered model} Flat bands are known
to host ferromagnetic phases in presence of repulsive interactions
\cite{Mielke1991,Mielke1991a,Mielke1992,Tasaki1992,Mielke1993,Tasaki1996}. The
presence of a gap to the flat band in our model ensures that the many-body
groundstate at filling $n=1/6$ is the unique fully-saturated ferromagnetic
state.

To see this in our model of disordered flat bands, we consider a fermionic
version with repulsive Hubbard interactions,
\begin{equation}
  \mathcal{H} =  \sum_{\left<i,j\right>,\sigma} \left(  t_{ij} \hat{c}^{\dagger}_{i\sigma} \hat{c}_{j\sigma} +c.c. \right)+  \sum_i \mu_i \hat{n}_i  + U \sum_i n_{\uparrow} n_{\downarrow}\, ,
  \label{eq:Hubbard}
\end{equation}
for spin 1/2 fermions, $n_i = n_{i\uparrow}+ n_{i\downarrow}$, $t_{ij}$ and
$\mu_i$ are chosen as above, and we consider the BDM to have a flat gapped
non-interacting band.

Since for $U>0$ the interaction term is positive, and the kinetic energy is
positive-definite by construction, many body states with $E=0$ are necessarily
groundstates.

One groundstate is easily obtained by filling the non-interacting flat band
completely with polarised spins which do not interact. Thus, we have at filling
$n = 1/6$ a ferromagnetic groundstate with maximal spin $S=L^2/2$, with the full
$2S (2S+1)$ degeneracy due to the $SU(2)$ symmetry of the model. The main
question to obtain ferromagnetism is whether this groundstate is unique, or if
there are additional non-magnetic states as well. Here, it turns out that the
groundstate is gapped, since the non-interacting band-structure has a finite gap
for the BDM.

We performed exact diagonalisation of the Hubbard model, Eq.~\ref{eq:Hubbard},
on small finite-size Kagome clusters ($2 \times 2$, $2\times 3$) to confirm that
the groundstate is indeed of the described form.

Finally, due to the presence of a spectral gap, we expect the ferromagnetism to
be stable to finite perturbations and fluctuations in the particle number.
Indeed, ferromagnetism is expected to be enhanced compared to the usual Kagome
case, since the localised non-interacting states now contain two hexagons.

\section{Outlook}
Demonstrating that the flat bands of the kagome lattice can be gapped opens up
the kagome lattice as a prime platform for the clean, i.e. isolated from the
dispersive bands by an arbitrarily large gap, study of topological and more
general flat band phenomena.

In addition, the presence of a flat band in a disordered model is highly
non-trivial and of general interest even if it requires fine-tuning between the
hopping and site-potential terms.

In terms of realisations of the specific type of couplings: We recall that this
model is naturally realised in the large-N limit
\cite{Stanley_1968,Garanin_1999} of a classical nearest-neighbour
bond-disordered Heisenberg-(Anti)ferromagnet, where the correlation between
site- and bond-disorder arrises from the spin length constraint. In other
settings it is unlikely that bond- and site-disorder is correlated in the
required way, thus, the system would need to be specifically designed. In this
case we envision it would be considerably easier to realise the translationally
invariant model reducing the required number of parameters that have to be
tuned. (For the minimal model we would require tuning 1 site-potential and 2
tunneling couplings in each unit cell). This might be feasible in cold-gas
setups where control over individual sites and bonds is possible by the use of
quantum gas microscopes.

In terms of topological properties of the flat band, we note that fluxes in the
MCM model are trivial by construction (since they can be removed by a unitary
gauge transformation). The BDM model in contrast supports non-trivial fluxes
along the hexagon loops of the lattice. However, since in the BDM model all
states of the flat band can be chosen localised, the non-interacting model is
necessarily topologically trivial \cite{Read2017}.

Our model also presents a natural realisation of flat band ferromagnetism on the
Kagome lattice, where the gap of the single particle spectrum results in a
unique gapped fully saturated ferromagnetic many-body state in presence of
repulsive on-site interactions. We reserve the further discussion of interacting
many-body phases in the gapped flat band and the effects on the magnon bands of
magnets for future work.

It might also be interesting to explore the effect of longer-range interactions
on the flat bands of this model which have recently been found to be remarkably
stable for the non-disordered model \cite{Maksymenko2017}.

\sectionn{Acknowledgements} We thank J. Richter for insightful discussions and
comments on a first draft of this paper. This work was in part supported by
Deutsche Forschungsgemeinschaft via SFB 1143.

%

\clearpage

\begin{center}
  \textbf{\large Supplemental Material:}
\end{center}
\setcounter{equation}{0} \setcounter{figure}{0} \setcounter{table}{0}
\setcounter{page}{1} \setcounter{section}{0} \makeatletter
\renewcommand{\thepage}{\arabic{page}}
\renewcommand{\thesection}{S\arabic{section}}
\renewcommand{\thetable}{S\arabic{table}}
\renewcommand{\thefigure}{S\arabic{figure}}
\renewcommand{\theequation}{S\arabic{equation}}
\renewcommand{\bibnumfmt}[1]{[S#1]} \renewcommand{\citenumfont}[1]{S#1}

\section{Construction of Flat Band States for BDM}
\begin{figure}
  \begin{minipage}{0.99\columnwidth}
    \includegraphics[width=.99\columnwidth]{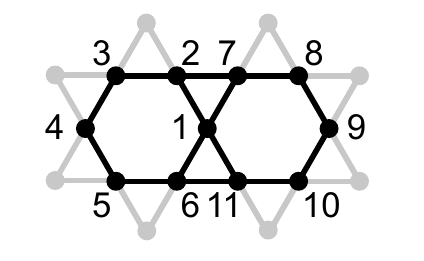}
  \end{minipage}
 \caption{A double hexagon of the kagome lattice. The wavefunction of a BDM zero
   energy state is localised on the black sites.
   Note that the state occupies 11 sites, and is part of 10 triangles, thus,
there are 11 degrees of freedom and 10 constraints, in addition to the
   wavefunction normalisation, implying that there is a unique solution for such
   a localised state. Same as Fig.~\ref{fig:illustration_double_hexagon} in main
   text, reproduced here to be self-contained.
   \label{fig:supp_illustration_double_hexagon}}
\end{figure}
In this section we provide additional detail on the explicit construction of the
zero-modes of the BDM model. The conventions are illustrated in
Fig.~\ref{fig:supp_illustration_double_hexagon}, same as in main text.

\subsection{Single Hexagon}
We begin by showing that for the BDM it is impossible to localise a state on a
single hexagon.

An intuitive picture of how states can be localised is as follows: Selecting a
subset of sites from the lattice, a necessary condition for a state with
amplitudes only on these sites is that the hoppings to sites outside the chosen
subset interfere destructively.

For a single hexagon loop, this requires that the tunnelling to all the points
of the ``star of david'' vanishes. Starting at say site 1, this then fixes all
the amplitudes of the wavefunction going around the loop step by step. However,
after going around the full loop, again arriving at site 1, we require that the
amplitude turns out to be the same we started with. This then sets a necessary
condition on the values of the couplings for single hexagon localised states to
exist.

Starting at say site 1, we require $\gamma_1^{\triangledown} \Psi_1 +
\gamma_2^{\triangledown} \Psi_2=0$, which we can solve for $\Psi_2 = -
\frac{\gamma_1^{\triangledown}}{\gamma_2^{\triangledown}} \Psi_1$, continuing
with $\gamma_2^{\vartriangle} \Psi_2 + \gamma_3^{\vartriangle} \Psi_3 = 0$,
which we solve for $\Psi_3 =
\frac{\gamma_1^{\triangledown}}{\gamma_2^{\triangledown}}
\frac{\gamma_2^{\vartriangle}}{\gamma_3^{\vartriangle}} \Psi_1 $, and similiarly
along the loop until we again arrive at $\Psi_1$.

Writing the final condition out explicitly yields
\begin{equation}
  \frac{\gamma_1^{\triangledown}}{\gamma_2^{\triangledown}}  \frac{\gamma_2^{\vartriangle}}{\gamma_3^{\vartriangle}} \frac{\gamma_3^{\triangledown}}{\gamma_4^{\triangledown}} \frac{\gamma_4^{\vartriangle}}{\gamma_5^{\vartriangle}}  \frac{\gamma_5^{\triangledown}}{\gamma_6^{\triangledown}} \frac{\gamma_6^{\vartriangle}}{\gamma_1^{\vartriangle}}  \Psi_1 = \Psi_1 \, ,
\end{equation}
or by rearrangement
\begin{equation}
  \frac{\gamma_1^{\vartriangle}}{\gamma_1^{\triangledown}} \frac{\gamma_3^{\vartriangle}}{\gamma_3^{\triangledown}} \frac{\gamma_5^{\vartriangle}}{\gamma_5^{\triangledown}}=\frac{\gamma_2^{\vartriangle}}{\gamma_2^{\triangledown}} \frac{\gamma_4^{\vartriangle}}{\gamma_4^{\triangledown}} \frac{\gamma_6^{\vartriangle}}{\gamma_6^{\triangledown}}
  \label{eq:star_condition}
\end{equation}
We note that this condition, in a slightly different notation and context, has
been derived previously \cite{Roychowdhury2017}.

Eq.~(\ref{eq:star_condition}) immediately shows that for the BDM for which
$\gamma^{\vartriangle} \neq \gamma^{\triangledown}$, e.g. for broken inversion
symmetry, single hexagon loops cannot exist. In the presence of inversion
symmetry, e.g. for the MCM, this condition is satisfied and states can be
localised on a single hexagon as is well known for the clean model.
\subsection{Double Hexagon}
Next we consider the double hexagon loop. As for the single hexagon we have some
sites on the periphery for which we require destructive interference (8
conditions), in addition we have two internal triangles (2 conditions), for in
total 10 conditions for 11 wavefunction amplitudes, which considering the
normalisation choice, can uniquely determine the wavefunction if a solution is
possible at all.

We have equations of the form
\begin{align}
  \gamma_2^{\vartriangle} \Psi_2 + \gamma_3^{\vartriangle} \Psi_3 &= 0 \quad \text{(periphery)} \\
  \gamma_1^{\triangledown} \Psi_1 + \gamma_2^{\triangledown}\Psi_2 +\gamma_7^{\triangledown} \Psi_7 & = 0 \quad \text{(internal triangle)} 
\end{align}

To determine the solubility of the resulting linear system of equations, one may
set $\Psi_1=1$ and consider the determinant of the matrix which turns out to be
\begin{multline}
  D = \gamma^{\triangledown}_3 \gamma^{\triangledown}_5 \gamma^{\triangledown}_7
  \gamma^{\triangledown}_9 \gamma^{\triangledown}_{11} \,
  \gamma^{\vartriangle}_2 \gamma^{\vartriangle}_4 \gamma^{\vartriangle}_6
  \gamma^{\vartriangle}_8 \gamma^{\vartriangle}_{10} - \\
  \gamma^{\vartriangle}_3 \gamma^{\vartriangle}_5 \gamma^{\vartriangle}_ 7
  \gamma^{\vartriangle}_9 \gamma^{\vartriangle}_{11} \, \gamma^{\triangledown}_2
  \gamma^{\triangledown}_4\gamma^{\triangledown}_6 \gamma^{\triangledown}_8
  \gamma^{\triangledown}_{10} \, .
\end{multline}
In the absence of inversion symmetry ($\gamma^{\vartriangle} \neq
\gamma^{\triangledown}$), this determinant is generically non-zero, and we
obtain a unique solution for the state localised on a double-hexagon.

We also note that in presence of inversion symmetry this determinant vanishes as
expected, since then we would have two linearly independent localised states on
each hexagon, and any linear combination of them would also form a state
localised on the double hexagon.
\section{Translationally invariant models}
In this section we present the explicit expression for the Hamiltonian in
momentum space, the flat band eigenstates at generic momentum and the gap at the
$\Gamma$-point for the BDM model discussed in the main text.

The Hamiltonian in momentum space reads
\begin{widetext}
  \begin{equation}
    \begin{pmatrix}
      \left| \gamma _A^{\triangledown }\right| ^2+\left| \gamma _A^{\vartriangle}\right| ^2 & \gamma _A^{\triangledown } e^{-i \vect{k} \cdot\vect{\delta} _{\mathrm{AB}}} \left(\gamma _B^{\triangledown }\right)^*+\gamma_A^{\vartriangle} e^{i \vect{k} \cdot\vect{\delta} _{\mathrm{AB}}} \left(\gamma _B^{\vartriangle}\right)^* & \gamma _A^{\triangledown } e^{-i \vect{k} \cdot\vect{\delta} _{\mathrm{AC}}} \left(\gamma _C^{\triangledown }\right)^*+\gamma _A^{\vartriangle} e^{i\vect{k} \cdot\vect{\delta} _{\mathrm{AC}}} \left(\gamma _C^{\vartriangle}\right)^* \\
      e^{i \vect{k} \cdot\vect{\delta} _{\mathrm{AB}}} \gamma _B^{\triangledown } \left(\gamma _A^{\triangledown }\right)^*+e^{-i \vect{k} \cdot\vect{\delta} _{\mathrm{AB}}} \gamma _B^{\vartriangle} \left(\gamma _A^{\vartriangle}\right)^* & \left| \gamma_B^{\triangledown }\right| ^2+\left| \gamma _B^{\vartriangle}\right| ^2 & \gamma _B^{\triangledown } e^{-i \vect{k} \cdot\vect{\delta} _{\mathrm{BC}}} \left(\gamma _C^{\triangledown }\right)^*+\gamma _B^{\vartriangle} e^{i\vect{k} \cdot\vect{\delta} _{\mathrm{BC}}} \left(\gamma _C^{\vartriangle}\right)^* \\
      e^{i \vect{k} \cdot\vect{\delta} _{\mathrm{AC}}} \gamma _C^{\triangledown
      } \left(\gamma _A^{\triangledown }\right)^*+e^{-i \vect{k}
        \cdot\vect{\delta} _{\mathrm{AC}}} \gamma _C^{\vartriangle} \left(\gamma
        _A^{\vartriangle}\right)^* & e^{i \vect{k} \cdot\vect{\delta}
        _{\mathrm{BC}}}\gamma _C^{\triangledown } \left(\gamma _B^{\triangledown
        }\right)^*+e^{-i \vect{k} \cdot\vect{\delta} _{\mathrm{BC}}} \gamma
      _C^{\vartriangle} \left(\gamma _B^{\vartriangle}\right)^* & \left| \gamma
        _C^{\triangledown }\right|
      ^2+\left| \gamma _C^{\vartriangle}\right|^2 \\
    \end{pmatrix}
  \end{equation}
\end{widetext}
with the site and triangle dependent couplings
$\gamma_i^{\vartriangle/\triangledown}$ which can generically be complex, the
momentum $\vect{k}$ and the difference vectors in the unit cell,
$\vect{\delta_{xy}} = \vect{r}_x - \vect{r}_y$

Since the expressions for complex couplings and general $\vect{k}$ get rather
unwieldy, we present them for real couplings only below.

The zero-mode at generic $\vect{k}$ is given by
\begin{widetext}
  \begin{equation}
    \left\{e^{-\frac{1}{4} i \left(k_1+\sqrt{3} k_2\right)} \left(\gamma_B^{\vartriangle} \gamma_C^{\triangledown} e^{\frac{1}{2} i \sqrt{3} k_2}-\gamma_C^{\vartriangle} \gamma_B^{\triangledown} e^{\frac{i k_2}{2}i}\right),e^{-i\frac{1}{4} \left(k_1+\sqrt{3} k_2\right)} \gamma_C^{\vartriangle} \gamma_A^{\triangledown}-\gamma_A^{\vartriangle} \gamma_C^{\triangledown} e^{\frac{1}{4} i \left(k_1+\sqrt{3} k_2\right)},  \gamma_A^{\vartriangle} \gamma_B^{\triangledown} e^{i k_1/2}- e^{-\frac{i k_1}{2}} \gamma_B^{\vartriangle}   \gamma_A^{\triangledown}\right\}
  \end{equation}
\end{widetext}
which explicitly shows that both models have a flat band at $E=0$.

These account for $N_s/3$ of the zero-modes. The remaining missing zero-mode for
the MCM is found at $\vect{k}=0$ where the dispersive band touches the flat band
which we discuss next.

Computing the eigenvalues at $\vect{k}=0$ allows us to see how the gap opens for
the BDM and remains closed for the MCM. These are given by $0$ and

\begin{widetext}
  \begin{equation}
    \frac{1}{2} \left(\Delta \pm \sqrt{\Delta^2-4 \gamma_{C}^{\triangledown 2}
        \left(\gamma_{A}^{\vartriangle 2}+\gamma_{B}^{\vartriangle 2}\right)-4
        \gamma_{B}^{\triangledown 2} \left(\gamma_{A}^{\vartriangle
            2}+\gamma_{C}^{\vartriangle 2}\right)-4 \gamma_{A}^{\triangledown 2}
        \left(\gamma_{B}^{\vartriangle 2}+\gamma_{C}^{\vartriangle 2}\right)+8
        \gamma_{C}^{\vartriangle} \gamma_{C}^{\triangledown} (\gamma_{A}^{\vartriangle}
        \gamma_{A}^{\triangledown}+\gamma_{B}^{\vartriangle}
        \gamma_{B}^{\triangledown})+8 \gamma_{A}^{\vartriangle}
        \gamma_{B}^{\vartriangle} \gamma_{A}^{\triangledown}
        \gamma_{B}^{\triangledown}}\right)
    \label{eq:supp_generic_eval}
  \end{equation}
\end{widetext}

with $\Delta =\gamma_A^{\vartriangle 2} +\gamma_A^{\triangledown 2} +
\gamma_B^{\vartriangle 2} +\gamma_B^{\triangledown 2}+ \gamma_C^{\vartriangle 2}
+\gamma_C^{\triangledown 2} $.

This manifestly shows that for the MCM where
$\gamma^{\vartriangle}=\gamma^{\triangledown}$ the second eigenvalue is also
zero corresponding to the band-touching point.

Further, for the BDM where lattice-inversion symmetry is broken, a gap is seen
to open up. Eq.~(\ref{eq:supp_generic_eval}) reduces to the expressions given in
the main text for the corresponding choices of couplings.

\section{Disordered Models}
Here we provide some additional details on the structure of the flat band states
in the disordered models which are not immediately apparent from our explicit
construction of all states.

\subsection{Structure of flat band states}
We next turn to analyse the structure of the new flat band states in the
disordered model.

To this end we consider the total weight of the dispersive states of the clean
model in the disordered flat band
\begin{equation}
  \left<\mathrm{FB}| \mathrm{NFB}_0  \right> = \sum_{\psi_0 \in \mathrm{NFB}}\frac{1}{N_{FB}}\sum_{\psi \in \mathrm{FB}} \left| \left< \psi | \psi_0\right> \right|^2
\end{equation}
and the corresponding state resolved-quantity, e.g. the projection of the
disordered flat band states $\psi$ onto the dispersive states of the clean model
$\psi_0$ averaged over the flat band
\begin{equation}
  \frac{1}{N_{FB}}\sum_{\psi \in \mathrm{FB}} \left| \left< \psi | \psi_0\right> \right|^2 \, .
\end{equation}

\begin{figure}
  \begin{minipage}{0.99\columnwidth}
    \includegraphics[width=.99\columnwidth]{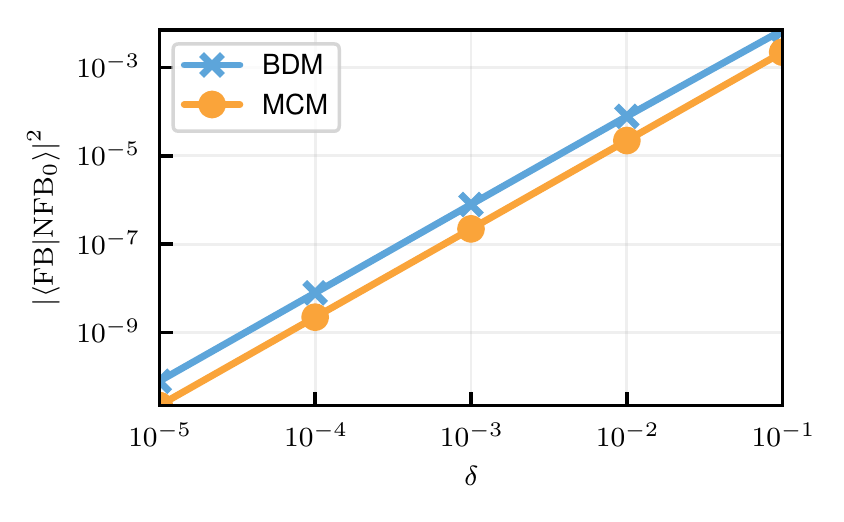}
  \end{minipage}

  \begin{minipage}{0.44\columnwidth}
    \includegraphics[width=.99\columnwidth]{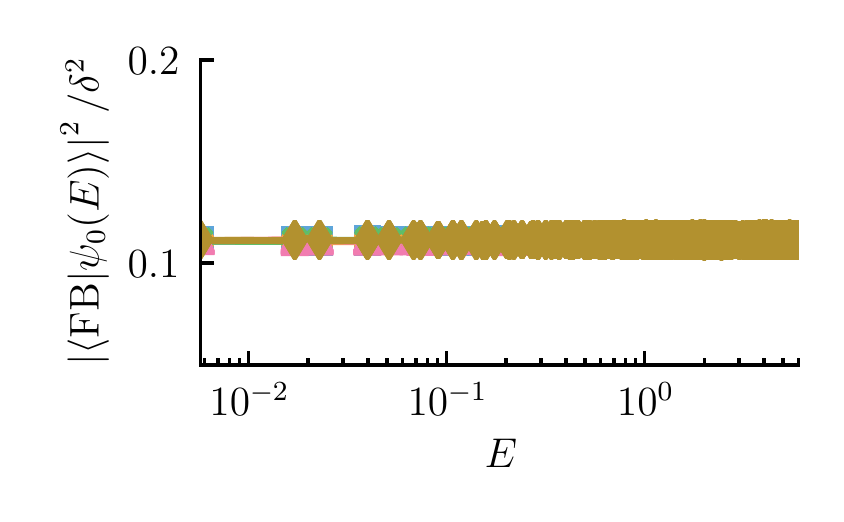}
  \end{minipage}
  \begin{minipage}{0.44\columnwidth}
    \includegraphics[width=.99\columnwidth]{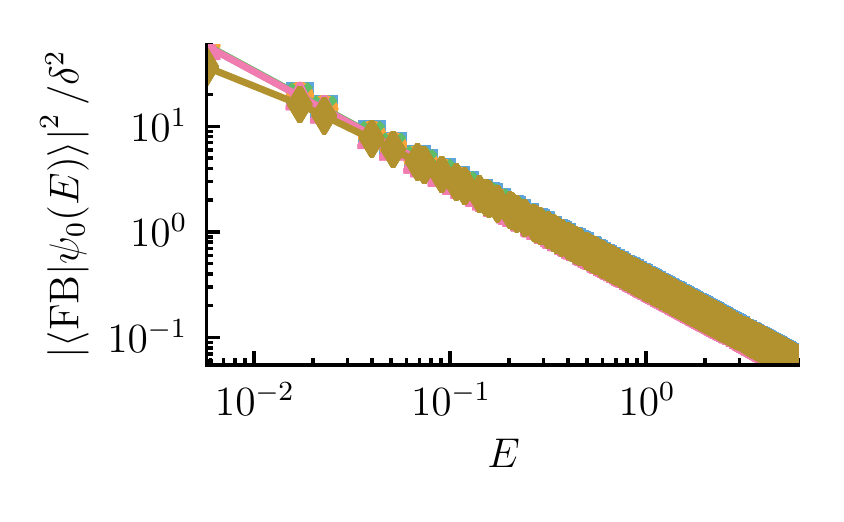}
  \end{minipage}
 \caption{ Top: Total weight of the dispersive states of the clean model of the
   flat band of the disordered model for MCM (circles) and BDM (x's)
   Bottom: Energy resolved weight of the dispersive states in the flat
   band states of the disordered model for MCM (left) and BDM (right), for
   $\delta=10^{-5},10^{-4},10^{-3},10^{-2},10^{-1}$ (squares,circles,x's,triangles,diamonds)
   normalised by the observed scaling $\sim \delta^2$. Note that the BDM is on a
   log-log scale and the MCM on a linear-log scale.
   \label{fig:weights}}
\end{figure}

The total weights and the state/energy resolved results for the MCM and BDM are
shown in the top and bottom of Fig.~\ref{fig:weights} respectively. For both
models we observe the same scaling $\left<\mathrm{FB}| \mathrm{NFB}_0 \right>
\sim \delta^2$ for the total weights.

However, the state/energy resolved weights differ qualitatively between BDM and
MCM. The BDM disordered flat band contains dominantly low-energy states of the
clean model and the squared amplitudes decay as a powerlaw $\sim 1/E$ with
increasing energy. In contrast, the MCM disordered flat band contains states of
all energies of the clean model with equal amplitudes independent of energy.

This can be traced to the fact that the matrixelements
$\expval{\mathrm{FB}}{\mathcal{H}}{\psi(E)}$ scale as $E^{1/2}$ and $E^1$ for
BDM and MCM respectively, which when combined with the expected $E^{-1}$ scaling
in second order perturbation theory explains the observed behaviour.

\end{document}